\documentclass[12pt]{article}
\usepackage{epsfig}

\renewcommand{\thesection}{\arabic{section}}

\catcode`@=11

\@addtoreset{equation}{section}
\@addtoreset{equation}{subsection}
\def\theequation{\ifnum\value{section}=0 \arabic{equation}\ignorespaces
\else \ifnum\value{section}=-1 A.\arabic{equation}\ignorespaces
\else \ifnum\value{subsection}=0 \thesection.\arabic{equation}\ignorespaces
\else \thesection.\arabic{subsection}.\arabic{equation}\ignorespaces
                             \fi
                        \fi
                   \fi}

{\catcode`\'=\active \def'{{}^\bgroup\prim@s}}

\catcode`@=12

\newcommand{\bq}{\begin{equation}}
\newcommand{\be}{\begin{equation}} 
\newcommand{\fq}{\end{equation}}
\newcommand{\ee}{\end{equation}}
\newcommand{\bqr}{\begin{eqnarray}}
\newcommand{\beqs}{\begin{eqnarray}} 
\newcommand{\fqr}{\end{eqnarray}}
\newcommand{\eeqs}{\end{eqnarray}}

\newcommand{\rf}[1]{(\ref{#1})}

\def\bop#1{\setbox0=\hbox{$#1M$}\mkern1.5mu
	\vbox{\hrule height0pt depth.04\ht0
	\hbox{\vrule width.04\ht0 height.9\ht0 \kern.9\ht0
	\vrule width.04\ht0}\hrule height.04\ht0}\mkern1.5mu}

\begin{document}
\thispagestyle{empty}

\begin{flushright} 
\begin{tabular}{l} 
UCLA-02-TEP-24\\
hep-th/0209072 \\ 
\end{tabular} 
\end{flushright}  

\vskip .6in 
\begin{center} 

{\Large\bf  Quantum Gravity with the Standard Model}

\vskip .3in 

{\bf Gordon Chalmers} 
\\[5mm] 
{\em Department of Physics and Astronomy \\ 
University of California at Los Angeles \\ 
Los Angeles, CA  90025-1547 } \\  

{e-mail: chalmers@physics.ucla.edu}  

\vskip .5in minus .2in 

\def\Prop{\Triangle}
\def\Prod{\prod} 

{\bf Abstract}   
\end{center}

A unfied theory of four-dimensional gravity together with the standard model 
is presented, with supersymmetry breaking of M-theory at a TeV.  Masses of 
the known particles are derived.  The cosmological constant is quantum 
generated to the observed value.  Quantum corrections to the classical 
compactification are analyzed, and the scenario is stable.

\setcounter{page}{0}
\newpage 
\setcounter{footnote}{0} 

\section{Introduction} 

Supersymmetry breaking at a TeV scale has emerged as the leading theoretical
candidate for the phenomenological formulation of standard model extensions.  

M-theory compactifications on $S^7$ manifolds \cite{Duff:1982ev,
D'Auria:1982mt} and Joyce manifolds \cite{Atiyah:2001qf,Acharya:2000gb} 
realize closely the generation of the standard 
model. In this paper we examine the formulation upon compactifying eleven  
dimensional \cite{Cremmer:1978km} M-theory \cite{Chalmers:M} 
on $dS_{3,1}\times S^7$, with twisting the $S^7$; there is a four-form 
flux turned on to make the background a solution to the classical 
equations of motion.  
Twisting the $R_{3,1}$ is also viable, as well as incorporating the 
Robertson-Walker spacetime in order to describe a different era in 
the cosmological evolution.  We consider these scenarios to model the 
spacetime with no inhomogeneties.\footnote{Perturbations 
of the underlying spacetime may be inserted to model large scale strucure.}  
The local spacetime is $R_{3,1}$, but there is a cosmological constant 
that is quantum generated; this constant agrees with data and modifies the 
geometry to de Sitter space.  String modes and branes are incorporated 
via wrapping membranes on the compact cycles of the internal geometry.  

In addition to the supersymmetry breaking and gauge group generation, the 
question of  the origin of mass is answered.  The known fermion masses may 
be obtained via a symmetry of the fermion interactions, and the masses come 
from perturbative resummations or membrane configurations in the compactified 
theory.  Breaking supersymmetry at approximately 2, or 1 to 3, TeV is ideal.  

The mass pattern is apparent and linear on a log scale and follow in 
accordance with a symmetry of the fermions; the associated charges are in
the centers of the $SO(3,1)$ and $SU(3)\times SU(2) \times U(1)$ gauge 
groups.  This fermionic symmetry requires the modded seven-sphere and 
is also made possible 
via modding the $R_{3,1}$.  In addition, there is a scenario at the GUT 
scale. 

The gauge couplings may be explained via running and unification up to the 
GUT scale, but we do not analyze this scenario in too much detail.  
The presented compactification scenario works in the GUT scale unified 
theory after adjusting some of the parameters.  The three couplings 
$\alpha_{QED}, \alpha_{EW}$, and $\alpha_{S}$ also have a symmetric 
form that comes about from the geometry of the compactified theory. 

The quantum modifications to the unified gravity/standard model are analyzed, 
and the corrections shown not to modify the theory, apart from the usual 
renormalization of the fields.  The setup is well-suited for supersymmetric 
phenomenology below a TeV, and collider energy scales are in the range 
to be tested by the Large Hadron Collider.  Gravity corrections are 
suppressed by the dimensionless parameter $\Lambda/m_{pl}$, with $\Lambda$ 
a TeV.  

In addition to model building, the feasibility of performing the complicated 
quantum calculations in gravity and gauge theory has improved much; there 
is progress in the art of gauge and gravity calculations, but little progress 
has 
been made beyond the multi-loop level due to the complications of the 
integrals.  Quantum chromodynamics amplitudes have been well studied to 
one-loop and partially to two loops using a variety of techniques.  
Additional 
progress has been in this direction through the use of the derivative 
expansion 
in three types of theories: scalar, maximally supersymmetric models, and 
non-supersymmetric gauge theories with matter.  This work may be 
examined in \cite{Chalmers:inprog}.  

The perturbative quantum structure of the gravity and standard model, 
in the derivative expansion is an infinite expansion 
in energies.  Loop graphs are conventionally an expansion in the coupling 
constant; the derivative expansion enables one to perform all loop 
integrations.  The theory becomes pseudo-classical after this reformulation,  
and the spectrum of quantum chromodynamics is analyzed.  The 
derivative calculations, presented in \cite{Chalmers:inprog}, allow for 
a non-perturbative 
definition of QCD together with a perturbative formulation of gravity.  This 
approach contains simple techniques to evaluate the quantum corrections in 
the gravitational theory, as well as the quantum corrections in the most 
general lagrangian formulation.

In subsequent work a holographic definition of QCD is given, containing 
gravity in one of the scenarios presented here.  Previous holographies 
equate gravity to gauge theory; this pseudo-holographic work has both 
states in the theory, without one being the bound state of another.  

The outline of this work is as follows.  In Section 2 we examine the 
fundamental constants of masses and gauge couplings.  Charges and 
symmetries are introduced, and the quantized masses are found from 
these charges.  In Section 3 we build the compactification scenarios 
containing the standard model and generate the masses and couplings 
of the standard model.  In section 4, the electroweak model and QCD 
are discussed in relation to the former.  In Section 4 we examine the 
quantum corrections to the gravity theory.  Cosmology and its 
pertinence to the model in this work is in section 5.  Discussions 
conclude. 

\section{Mass and Coupling quantizations}

The masses of the fermions, and bosons, are examined and derived via the 
geometry of the underlying gravity theory.  A feature of having geometry 
break the supersymmetry as well as generate the fermion masses that we 
do not require a chiral theory initially.  QCD is believed to have a chiral 
phase 
transition at $T=0$ in $d=4$ and this conjecture is not required; 
furthermore, 
anomaly cancellation is avoided and leaves more room for further generations.

The geometry, and contributing M2 and M5 branes, generate a symmetric 
mass pattern agrees with observed phenomenology of the standard model; in 
fact, this work via extrapolating the fermionic symmetry would of
predicted the top 
quark mass.  The masses are labeled in accord with the integers, 

\bqr 
\Lambda ({\Lambda\over m_{pl}})^{n_j/16} \equiv \Lambda \rho^{n_j/16}
\fqr  
and follow from the representations of $SU(3)/ Z_3 \cdot U(2)/Z_2 \cdot   
SO(3,1)$, the standard model gauge group in four dimensions.

The masses of the observed fermions fall into the pattern for the leptons, 
with masses measured in electron Volts,
\bqr
m_e = \alpha_e \Lambda \rho^{5/16} \quad\quad .5\times 10^{6} 
\fqr
\bqr
m_\mu = \alpha_\mu \Lambda \rho^{4/16} \quad\quad 1.1\times 10^8 
\fqr
\bqr
m_\tau = \alpha_\tau \Lambda \rho^{3/16} \quad\quad 1.7\times 10^9 
\fqr
and the quarks 
\bqr
m_t = \alpha_t \Lambda \rho^{1/16} \quad\quad 1.7\times 10^{11} 
\fqr 
\bqr
m_b = \alpha_b \Lambda \rho^{2/16} \quad\quad 4.7\times 10^9 
\fqr  
\bqr
m_c = \alpha_c \Lambda \rho^{3/16} \quad\quad 1.5\times 10^9
\fqr
\bqr
m_s = \alpha_s \Lambda \rho^{4/16} \quad\quad 1.5\times 10^8
\fqr 
\bqr
m_u = \alpha_u \Lambda \rho^{5/16} \quad\quad 8\times 10^6
\fqr
\bqr
m_d = \alpha_d \Lambda \rho^{6/16} \quad\quad 4\times 10^6 \ . 
\fqr
The parameters $\alpha$ are of order unity, and the $\rho$ variable is 

\bqr 
\rho=\Lambda/m_{pl} \quad\quad m_{pl}=2.1\times 10^{28} ~{\rm eV}\ .
\fqr 
The parameters are obtained from the classical breaking of 
supersymmetry, at the TeV 
scale.   All of these numbers, except two that miss be a factor of 2 and 3, 
match the fermions if we have $\Lambda=10^{12}$ ev; of course this is not a 
prediction for supersymmetry at a TeV, but it is suggestive in the $1-3$ 
range.  
The $u$ and $d$ quarks are almost degenerate and we may interchange 
them with the caveat that quantum corrections are the cause.  

It is important 
to note that chirality does not enter into the description of the fermions 
as they are massive, and neither does $SU(3)$ anomaly cancellation 
(although the content cancels the anomaly).  We
find the group theory origin 
of these quantum numbers in the following.  The pattern is quantization of 
the 
masses, and there is a sum rule, 

\bqr 
\sum_l  n_l = {4\over 7} \sum_q n_q \ ,
\fqr 
from 
\bqr 
\sum_q m_q = 21, \qquad\quad \sum_l m_l = 12  \ .
\fqr 

The neutrino content is believed to be approximately .01 ev.  We 
split the three neutrinos as 
\bqr 
m_{\nu_e} = \alpha_{\nu_e} \Lambda \rho^{15/16}  \quad\quad .001 
\fqr 
\bqr 
m_{\nu_{\mu}} = \alpha_{\nu_\mu} \Lambda \rho^{14/16}  \quad\quad .01 
\fqr 
\bqr 
m_{\nu_{\tau}} = \alpha_{\nu_\tau} \Lambda \rho^{13/16}  \quad\quad .1  \ , 
\fqr 
The netrinos may also be generated via a perturbative gravity calculation, 
\bqr 
\alpha {\Lambda^2\over m_{pl} } = .01 ~{\rm eV} \ ,  
\fqr 
with $\alpha\sim 10$ from the one-loop factors.  This corresponds to a 
one-loop gravity calculation.  The sum rule for these neutrinos is 
\bqr 
\sum_{\nu} m_{\nu} = 2 \sum_q m_q = 7/2 \sum_l m_l \ , 
\fqr 
with, 
\bqr 
\sum_{\nu} m_\nu = 42 \ .  
\fqr 
Neutrinos dont couple to photons, and barring the SU(3) couplings, this 
one-loop gravity calculation would generate the neutrino masses; the 
higher-loop contributions in gravity mixed with matter are suppressed by 
$\Lambda/m_{pl}= 10^{-16}$.  This is the second mechanism for generating 
the observed values of the neutrino masses, without the mass splittings to 
first order.  

The gauge bosons have masses 
\bqr 
m_{W^\pm} = \Lambda \rho^{1/16} \quad\quad 8\times 10^{10} 
\fqr 
\bqr 
m_{Z} = \Lambda \rho^{1/16} \quad\quad 9.2\times 10^{10} \ . 
\fqr 
\bqr 
m_{\gamma}=m_{{\rm glue}}=m_{{\rm graviton}} = 0  
\fqr 
Scalars generically get perturbative masses on the order of a TeV; if the 
mass were generated gravitationally then it would be in the 100 GeV range.  

The remaining scales are:

\bqr 
\Lambda_{QCD} = \Lambda \left( {\Lambda\over m_{pl}}\right)^{6/16} 
\fqr 
\bqr 
\Lambda_{EW} = \Lambda \left( {\Lambda\over m_{pl}}\right)^{1/16}  \ .
\fqr 
Breaking the supersymmetry at TeV scale generates the observed mass 
pattern of the standard model, and the quantum gravitational sector is 
incorporated in a direct fashion.  

The gauge couplings at the strong scale point are 
\bqr  
\alpha_{QED} = {1\over 137} \sim {1\over (23) 6} \ ,
\fqr   
\bqr 
\alpha_{EW} = {1\over (23) 3} \left({2\over 3}\right)^{1/2} \ ,
\fqr 
\bqr 
\alpha_S \sim 1 \ ,
\fqr 
with the coupling relation, 
\bqr 
\alpha_{QED} = \left({8\over 3}\right)^{1/2} \alpha_{EW}  \ .
\fqr 
These factors may be interpreted via the underlying geometry.  This 
completes the review of the masses and couplings of the standard model, 
with the exception of the Higgs mass.  

The final number we examine is the cosmological constant; the experimental 
value is 
\bqr 
\Lambda^4 \left( {\Lambda\over m_{pl}}\right)^4 \ ,   
\label{cosmo}
\fqr 
with $\Lambda$ at 2 TeV.  This number is derivable if we break a maximally 
supersymmetric gravity theory in such a way that there are eight independent 
scales.  The breaking pattern is generated by this work, and as such, the 
value of the cosmological constant in this era is generated.  The 
corrections are ${\cal O}(\Lambda/m_{\rm pl})$ and are suppressed.  
We turn to deriving the mass quantizations.  

\section{Geometry of Masses} 

The following model is contained in M-theory, with supersymmetry breaking 
at a around a TeV (LHC).  We begin with a $S^7/Gamma \times R_{3,1}'$ 
background in eleven dimensions; in subsequence we examine a $S^7\times 
R_{3,1}/\Gamma$ to be used in a holographic formulation of QCD.   The seven 
sphere in the first case contains four independent parameters, 

\bqr 
\lambda z_i^a {\bar z}_{i,a}  + \lambda_7 X_7^2 +  
\lambda  X_8^2 = R^2_7 \ , 
\fqr 
with the $a$ index labeling the components of the complex number 
$z_i$.  To begin we take a quotient $X_7\leftrightarrow -X_7$ and 
$X_8\leftrightarrow -X_8$ to eliminate the extraneous fields generating 
along these cycles, and denote this by $\Gamma_2$.  The geometry 
contains an $SU(3)\times SU(2)$ (the index $a=1,2,3$) group of isometries.  
One may think of this as coming from an $SU(5)$ group,  via a complexified 
SO(5), from the metric 

\bqr 
\lambda z_i z_i + {\bar\lambda} {\bar z}^i {\bar z}^i+ \lambda_7 X_7^2 +  
\lambda  X_8^2 = R^2_{S_7} \ . 
\fqr 
The $R_{3,1}$ term with a quantum induced constant becomes $dS^4$.  
We begin with the topology, 

\bqr sum_{j=1}^4 \lambda_j X_j^2 = R^2_{dS}  \ ,  
\label{desitter} 
\fqr 
possessing five independent parameters, and a weak curvature.  The 
quantum gravity with eight parameters of supersymmetry breaking generates 
the value in eqn. (\rf{cosmo}) as a two-loop effect.  We work with the 
geometry in eqn. \rf{desitter} stabilized around this value, which is almost 
the flat $R_{3,1}$.  This is the simplest case, and it is easy to analyze 
the early universe with a supermassive black hole and the Robertson-Walker 
geometry.   

In order to achieve the mass pattern we require a symmetry of the fermions 
such that only the gravitational instantonic terms, and their quantum 
corrections, 
are allowed to contribute to the particle's masses.  This could be achieved 
in 
conjuction with a monodromy of the quantum theory with respect to the 
supersymmetry breaking scale, $\Lambda \rightarrow e^{2\pi i} \Lambda$. 

The Lagrangian for the four components of fermions is, 

\bqr 
{\cal L} = {\bar\psi}^{\dot\alpha} i\partial_{\dot\alpha}^\alpha \psi_\alpha 
+  {\bar\chi}^{\dot\alpha} i\partial_{\dot\alpha}^\alpha \chi_\alpha 
+ m\left(\psi^\alpha \chi_\alpha + 
 {\bar\psi}^{\dot\alpha} {\bar\chi}_{\dot\alpha}\right)  
\fqr 
The simplest symmetry is 

\bqr 
\psi_\alpha^a \rightarrow e^{-2\pi i n/30} \psi_\alpha^a \qquad
\chi_\alpha^a \rightarrow e^{-2\pi i n/30} \chi_\alpha^a 
\fqr 
and 

\bqr 
\bar\psi_{\dot\alpha}^a \rightarrow e^{2\pi i n/30} 
 \bar\psi_{\dot\alpha}^a \qquad 
\bar\chi_{\dot\alpha}^a \rightarrow e^{2\pi i n/30} 
 \bar\chi_{\dot\alpha}^a \ , 
\fqr 
combined with the mass term rotating due to the monodromy, 
\bqr 
m\rightarrow m ~ e^{2\pi i n/15} \ . 
\fqr 
In four-component notation we have, 

\bqr 
\psi \rightarrow e^{- 2\pi i n/30} \psi\ , \quad   
\bar\psi \rightarrow e^{- 2\pi i n/30} \bar\psi \ . 
\fqr 
With these transformations the mass terms for the fermions are 
invariant when, 

\bqr 
\Lambda ({\Lambda\over m_{pl}})^{n/15} \bar\psi \psi , 
\fqr 
which is the monodromy of $\Lambda \rightarrow e^{2\pi i} \Lambda$.  
The charge assignments for the fermions follows from the compactification on 
the $S^7/\Gamma$ and its quotient.  

We list the appropriate quotient $\Gamma$ of the $S^7$.  The masses are 
generated through the symmetry generated by the twist of the $S^7$ 
coordinates:   

\bqr 
\pmatrix{ z_1   &    -1=e^{\pi i} &   & & z_1 \cr  
              z_2 \rightarrow    &    & 1               &   & z_2 \cr 
              z_3   &    &  & -1=e^{-\pi i}  & z_3 } 
\fqr 
\bqr  
\pmatrix{ z^1   &  -1=e^{5i \pi}  & & z^1 \cr 
              z^2    &  & 1=e^{-2\pi i} & z^2 } \ , 
\fqr 
with the six coordinates labeled as $z_j^a$ (pertaining to the SU(2) and 
SU(3)); this action is $G_a$.  

The second action is, 

\bqr 
\pmatrix{ z_1   &    -1=e^{\pi i} &   & & z_1 \cr  
              z_2 \rightarrow    &    & 1               &   & z_2 \cr 
              z_3   &    &  & -1=e^{-\pi i}  & z_3 } 
\fqr  
and, 
\bqr  
\pmatrix{ z^1   &  1=e^{2i \pi}  & & z^1 \cr 
              z^2   &  & -1=e^{7\pi i} & z^2 } \ , 
\fqr 
which we label $G_b$.  These $Z_3\times Z_2$ actions are in the center of  
$SU(2) 
\times SU(3)$ and have a $U(1)$ component.  The $Z_3$ pertains to $z_i$ 
picking one of three factors, and the $Z_2$ to the two values given a single 
$z_i$ coordinate.    There are six possibilities of the quotient, three 
$z_i$ and two $z^a$; this corresponds to the sextet structure, or the 
doublet-triplet, of all of the fermions.  We give the charges below.  
The commutation of $G_a$ and $G_b$ requires a $Z_3$ action.   

The symmetry of the $S^7$ quotient with the generators above affect the  
fields in $R_{3,1}$ via the compactification.  The net quantum 
four-dimensional 
field $\psi$ is $\psi_\psi(X) \times \phi^a(Y)$, with $\phi^a$ carrying the 
$SU(3)/ Z_3\times SU(2)/Z_2 \times U(1)$ indices.  The $X$ and $Y$ are 
coordinates of the base and compactified spaces.  The group action on the 
compact fields generates a projection, by definition,   

\bqr 
\phi(G\cdot Y)= G\cdot \phi(Y) .  
\fqr 
In this manner there is an action on $\psi$, the phase because the group 
pertaining to the quotient action is diagonal,  so that the net field 
$\phi(Y) 
\times \psi(X)$ is invariant.  The kinetic term, 

\bqr 
\int d^4x\sqrt{g} \bar\psi\partial\psi
\fqr 
is invariant, and the four-dimensional symmetry allows a mass term only 
if there is a compensating transformation on the mass.  This is via the 
monodromy of $\Lambda$, the supersymmetry breaking scale.  

The fermion assignments due the group action follow from the quotient, 
or rather from the center of the gauge group $SU(2)\times SU(3)$ together 
with the $U(1)$.  The phases are for the sextet of the quarks ,  

\bqr 
\pmatrix{ SU(3)  \cr 
             1    \cr  
             0     \cr 
             -1  \cr }  
\qquad\quad 
\pmatrix{  SU(2)\times U(1)  \cr 
                   5 \cr 
                   2 } 
\fqr  
with six actions, 

\bqr 
\pmatrix{  
t & 6 \cr 
b & 5 \cr 
c & 4 \cr 
s & 3 \cr 
d & 2 \cr 
u & 1 } \ . 
\fqr 
The same for the leptons and quark 

\bqr 
\pmatrix{ SU(3)  \cr 
             1    \cr  
             0     \cr 
             -1  \cr }   
\qquad\quad 
\pmatrix{  SU(2)\times U(1)  \cr 
                   4 \cr 
                   14 } 
\fqr  
with six actions, 

\bqr 
\pmatrix{  
e & 5 \cr 
\nu_e & 15 \cr 
\mu & 4 \cr 
\nu_\mu & 14 \cr 
\tau & 3 \cr 
\tau_{\nu_\tau} & 13 } \ . 
\fqr 
The fermions transform in the way described, and 
these combinations allow the classical mass term exhibited earlier.  
There are corrections, of course, allowed by the symmetry, which include 
both perturbative and non-perturbative terms; the former comes from the 
usual loop graphs and the latter with M2 and M5 branes wrapped on the 
$S^7$ for example.  

The charge assignments follow naturally with the doublet-triplet structure 
of the fermions.  The fermion doublets and triplets are 

\bqr 
\pmatrix{ d \cr s \cr b } \qquad\quad \pmatrix{ u \cr c \cr t}  
\fqr 
and,  

\bqr  
\pmatrix{ e \cr \nu_e} \qquad\quad  \pmatrix{ \mu \cr \nu_\mu} \qquad\quad 
\pmatrix{ \tau \cr \nu_\tau} \ . 
\fqr  
The U(1) charge assignment for the quarks and lepton/neutrino doublets 
are $(2, -1)/3$ and $(-1, 0)$.  One may think of the $\pm 1$ in the SU(3) 
as $3({1\over 3})$.  

An alternate charge assignment, in conjuction with a mass displacement, 
potentially due to a singularity is the following.  Flip the SU(3) assignment 
to 
$(-1,0,1)$.  One may extract the $(5, 2)$ and $(4, 14)$ to, $(2 -1) + 3$ and 
$(-10, 0) + 14$; the two factors are a $U(1)$ charge and a uniform constant 
for all of the fermions.  The charge assignments then have the interpretation 
of a $SU(3)$ charge, $U(1)$ charge, and a constant.  The constant may have 
the interpretation of the contribution to the mass from a resolved 
singularity.  (Another phase split is to write $(5,2)=(3/2,-3/2)+7/2$ and 
$(4, 14)=(3/2, 21/2)+ 7/2$.)  This procedure can be carried through on 
$dS_{3,1}\times S^7$ with a quotient in and singularity of the $dS_{3,1}$; 
in this case the interpretation is that of $N=8$ supergravity with a 
singularity, and resoluton to a black hole.

\section{QCD and Electroweak sector} 

The QCD and Electroweak sector follow in a direct sense from the 
previous calculations.  There are scalars arising from the $7$ and $8$ 
directions of the $S^7$ and the matter content for the gauge theory is 
already present.  The Lagrangian for the gauge sector is, 

\bqr 
{\cal L}_{g} = \int d^4x \sqrt{g} 
~\left({\rm Tr} F^2_{SU(3)} + {\rm Tr} F^2_{SU(2)} -{1\over 4} F^2_{U(1)} 
\right)   
\fqr 
which is generated from the $S^7$ because the group action is taken 
in the center.  The action for the gravititational part is, 

\bqr 
{\cal L}_{gr}  = \int d^4 \sqrt{g} ~ \left(R + V\right) \ , 
\fqr 
and the fermionic components are, 

\bqr  
{\cal L}_f = \sum_j \int d^4x \sqrt{g} ~ \left( {\bar\psi}_j (i\partial - 
\omega) \psi_j \right) \ , 
\fqr  
with $\omega$ the total spin connection from the gauge degrees of freedom.  

The Higgs is required for the conventional mass generation of the guage 
bosons and fermions;  the scalars present may generate the masses for the 
$SU(2)$ and arise from the fields on the $z_i^a$ sector of the $S^7/\Gamma$. 

\bqr 
{\cal L} = \int d^4x\sqrt{g}~ \left( \nabla_{U(2)_\mu}\phi \nabla_{U(2)}^\mu 
{\bar\phi} 
   + {1\over 2} m^2 \phi{\bar\phi}\right)  + \lambda_\phi \phi^4 + \ldots
\fqr 
the remaining sector of scalars do not couple directly to the U(2) and 
the potential term is due to the non-supersymmetry of the background. 
The mass term is generically at the TeV scale due to quantum effects to 
the scalar propagator.  The minimal coupling to the gauge field is 
sufficient to generate the mass term for the SU(2) gauge fields.  
The gamma$_5$ couplings have not been examined in detail.

There are eight scalar modes in N=8 supersymmetric theory on 
$AdS_4\times S^7$ with all radii different.  As there are three couplings 
dictating the shape of the $S^7$, there are four massless moduli describing 
the zero mode perturbations.  Only one complex scalar is charged under the 
$U(2)$ and the other two, pertaining to the seven and eighth direction 
do not couple;  this combination is sufficient to generate the masses of 
the W's and Z's through the standard Higgs mechanism.  The scalar content 
is in agreement with the MSSM.  There are also towers of KK modes that 
may couple, and their masses are larger than a TeV.

\section{Quantum Corrections} 

The masses and low-energy amplitudes are stable under quantum
corrections.  In detail to one loop the matter fields do not correct the 
mass forms as only the Z-factors are relevant to renormalize to this 
order; in general to higher loops the matter and gravitational degrees 
of freedom do not alter the masses a substantial amount, but might 
explain the $d$ and $u$-quark splittings.  The scenario is preserved 
under the logs and power corrections.  

There are two types of quantum corrections to the background and 
fields: log corrections that modify the exponentiated term, 

\bqr 
\Lambda \Bigl[ \left({\lambda\over m_{\rm pl}}\right)^{n/16} \Bigr]^m 
\times \left( \ln\Lambda, \ln\ln\Lambda, \ln^p\Lambda\right)  \ , 
\fqr 
and combinations.  Second, we have further exponentiated terms that 
arise from wrapping M2 and M5 branes on the compact space, and 
other non-perturbative corrections in the $dS_{3,1}$; these generate 

\bqr 
\Lambda \Bigl[ \left({\lambda\over m_{\rm pl}}\right)^{n/16} \Bigr]^m \ , 
\fqr 
These terms may arise in a manner that respects the symmetry of 
the fermionic kinetic terms.  Usual renormalization theory takes care 
of large corrections in the Z-factor or other parameters in the standard 
model.  The $\Lambda \rightarrow \Lambda e^{2\pi i}$ symmetry is 
restrictive in the same sense as the $\Gamma_5$ chiral structure 
prohibits masses in chiral theories.  Note also that log terms may 
exponentiate via, 

\bqr 
x^{n/16} = e^{n/16 \ln x} = \sum_{j=1}^\infty \left( {n\over 16} \ln 
x\right)^j \ . 
\fqr 
Such a term appears order by order to violate this symmetry, but the 
infinite summation is in accord.  Perturbation theory must obey the 
symmetry, and to one loop order it does without summing.  

Most of the corrections from gravity are minor, being suppressed 
by $\Lambda/m_{pl}$.  However, because the neutrinos are so light, 
the question is raised if a gravity correction might modify its mass 
substantially.  A one-loop contribution to the fermions masses, which 
is combatible with the fermion symmetry, is 

\bqr 
\beta {\Lambda^2\over m_{pl}} = .0001 \beta \ , 
\fqr 
and is a factor smaller than the lightest neutrino mass in this scenario.  
Of course there is a factor $\beta$ from the loop integral but to first 
approximation 
this result is smaller, or on the order of the lightest candidate neutrino 
mass.  

The gravity sector is trivial because of the $\Lambda/m_{pl}$ factor; 
being an effective theory up to a TeV scale, in a string-inspired 
regulator through the M-theory formulation, the gravitational corrections 
are a power series, 

\bqr 
A_n = \sum_{j=0}^\infty A^{(j)}_n \left({\lambda\over m_{\rm pl}}\right)^j 
\fqr  
with $\Lambda/m_{\rm pl}= 10^{-16}$, a very small parameter. Expansions 
in loops (or $h/2\pi$) is the most traditional means to generating 
the perturbative expansion describing amplitudes in a quantum field or 
string theory.  However, such expansions are typically problematic in 
that multi-loop diagrams are difficult to compute (in the exception of 
ladder diagrams or certain infinite subsets of diagrams).  In 
\cite{Chalmers:M} the initiation of derivative 
expansions in the context of gravity theories was initially explored; the 
derivative expansion for QCD, scalar, and $N=4$ supersymmetric gauge 
theories was recently developed and the same techniques are useful in 
the perturbative gravitational context.  Integrals may be eliminated and 
perturbative gravitational amplitudes may be constructed through a recursive 
and algebraic approach.  We refer the reader to these papers. 

\section{Cosmology} 

The model here has a cosmological constant, 

\bqr 
\int d^4x \sqrt{g} V(\Lambda,m_{pl}), \qquad\quad 
V(\Lambda,m_{pl}) = \Lambda^4 \sum_{n=2}^\infty 
\left({\Lambda\over m_{pl}}\right)^{2n} \ . 
\fqr 
The first term is, with $\Lambda=1$ TeV, $10^{-15}$ erg/cm${}^3$ and is in 
agreement with cosmological data.  The curvature of space induced by this 
constant is 

\bqr 
R_{dS}^2 = {1\over \Lambda^2} \left({\Lambda\over m_{pl}}\right)^{-4}   
\fqr 
and is roughly 15 billion light-years.  This value of the cosmological 
constant, and its small corrections, are in accord with the large scale 
structure of the universe.  

The de Sitter space solution may be regarded at late times from the big bang, 
in models of the early universe.  We may replace the de Sitter space with the 
Robertson-Walker topology.  Its metric is, 

\bqr 
ds^2 = -dt^2 + a^2(t) \bigl[ {1\over 1+kr^2} dr^2 + r^2 
\left(d\theta^2+sin^2\theta 
d\phi^2\right) \bigr] \ . 
\fqr  
The warp factor is, 

\bqr 
H_0 = {{\dot a}\over a} \vert_0  \qquad q_0 = -H_0^{-2} \left({\dot{\dot 
a}\over a}\right)  
\fqr  
in terms of present day Hubble and deceleration parameters.  At late times, 
$a^2(t) \rightarrow 1$, and the spacetime is flat.  A modification of the 
Robertson-Walker metric is 

\bqr 
ds^2 = -\lambda_t dt^2 + a^2(t) \left( \lambda_{i} dx_i dx_i \right) \ , 
\fqr 
and contains four free-parameters, pertinent to breaking supersymmetry with 
the maximal number of free parameters.  At late times, this metric may be 
replaced with the de Sitter space soluton considered in this paper, 

\bqr  
\sum_{j=1}^4 \lambda X_i^2 = R^2_{dS} \ . 
\fqr  
Perturbations due to massive objects are expected to change this 
idealic geometry.

\section{Discussion}

The standard model content, masses and couplings, have been examined 
in detail in the context of M-theory compactified on a $S^7$.  The 
masses arise from classical gravitational configurations, from 
M2 and M5 branes, and the couplings from the geometry of the $S^7$.  

The masses of the standard model are quantized and fit experimental 
data.  The quarks and leptons are have these masses in line with the 
high energy symmetries of M-theory and/or supergravity.  The origin 
of mass is gravitational from this point of view, and they are 
derived in this work.  The couplings suggest grand unification at 
$10^{24}$ eV, but this is not required.  

Supersymmetry breaking in the model considered here occurs at 2 TeV 
with some flexibility.  With this breaking the cosmological constant 
comes out to the experimental value also.  The quantum corrections, 
including gravitational, are examined and do not substantially alter 
the predictions made at the classical level.  In addition to the 
M-theory on $S^7$, there is a pure 4-D background solution leading 
to the standard model parameters.

\section*{Acknowledgements} 

GC thanks the DOD, 444025-HL-25619, for support, and Emil Martinec 
for correspondence.


\begin{thebibliography}{99}  

\bibitem{Duff:1982ev}
M.~J.~Duff,
Nucl.\ Phys.\ B {\bf 219}, 389 (1983).

\bibitem{D'Auria:1982mt}
R.~D'Auria, P.~Fre and P.~van Nieuwenhuizen,
Phys.\ Lett.\ B {\bf 122}, 225 (1983).

\bibitem{Atiyah:2001qf}
M.~Atiyah and E.~Witten,
arXiv:hep-th/0107177.

\bibitem{Acharya:2000gb}
B.~S.~Acharya,
arXiv:hep-th/0011089.

\bibitem{Cremmer:1978km}
E.~Cremmer, B.~Julia and J.~Scherk,
Phys.\ Lett.\ B {\bf 76}, 409 (1978).

\bibitem{Chalmers:M}
G.~Chalmers,
Phys.\ Rev.\ D {\bf 64}, 046014 (2001)
[arXiv:hep-th/0104132].

\bibitem{Chalmers:inprog}
G.~Chalmers,
Scalar field theory in the derivative expansion; Gauge theories 
in the derivative expansion; $N=4$ supersymmetric gauge theories in 
the derivative expansion, to appear.

\end{thebibliography}
\end{document}